\documentclass[review]{elsarticle}

\makeatletter
\def\ps@pprintTitle{%
 \let\@oddhead\@empty
 \let\@evenhead\@empty
 \def\@oddfoot{}%
 \let\@evenfoot\@oddfoot}
\makeatother
\usepackage{ifpdf}
\usepackage{threeparttable}
\usepackage{subfigure}
\usepackage{graphicx}
\usepackage{graphics}
\usepackage{amsmath}
\usepackage{amssymb}
\usepackage{amsfonts}
\usepackage{mathtools}

\usepackage{stfloats}
\usepackage{float}
\usepackage{lipsum}
\usepackage{multirow}

\usepackage{color} 

\restylefloat{figure}

\usepackage{enumitem}
\usepackage{algorithm}
\usepackage{algorithmic}

\usepackage{tikz}
\usetikzlibrary{shapes.geometric, arrows}

\journal{Reliability Engineering \& System Safety}

\begin{document}

\begin{frontmatter}

\title{\Large{Infrastructure Recovery Curve Estimation Using Gaussian Process Regression on Expert Elicited Data}}

\author[mymainaddress]{Quoc~Dung~Cao}

\author[mysecondaryaddress]{Scott B. Miles}
\ead{milessb@uw.edu}
\author[mymainaddress]{Youngjun Choe\corref{mycorrespondingauthor}}
\cortext[mycorrespondingauthor]{Corresponding author}
\ead{ychoe@uw.edu}

\address[mymainaddress]{Department of Industrial \& Systems Engineering, University of Washington, Seattle}
\address[mysecondaryaddress]{Department of Human Centered Design \& Engineering, University of Washington, Seattle}



\begin{abstract}
Infrastructure recovery time estimation is critical to disaster management and planning. 
Inspired by recent resilience planning initiatives, we consider a situation where experts are asked to estimate the time for different infrastructure systems to recover to certain functionality levels after a scenario hazard event. We propose a methodological framework to use expert-elicited data to estimate the expected recovery time curve of a particular infrastructure system. This framework uses the Gaussian process regression (GPR) to capture the experts’ estimation-uncertainty and satisfy known physical constraints of recovery processes. The framework is designed to find a balance between the data collection cost of expert elicitation and the prediction accuracy of GPR. We evaluate the framework on realistically simulated expert-elicited data concerning two case study events, the 1995 Great Hanshin-Awaji Earthquake and the 2011 Great East Japan Earthquake. 
\end{abstract}


\end{frontmatter}

\normalsize
\section{Introduction}
This work is motivated by a recent trend of resilience planning initiatives. Our current ability to estimate infrastructure recovery trajectories is limited, as revealed in the recent resilience planning efforts of U.S. communities, which started in San Francisco, CA \cite{spur2009resilient} and became state-wide initiatives in Washington State \cite{washington2012resilient} and Oregon \cite{osspac2013oregon}. These efforts inspired the National Institute of Standards and Technology (NIST)’s Community Resilience Planning Guide \cite{nist2016} as a model for other jurisdictions. The current estimation practice is largely ad hoc. 
Although there is a growing body of literature on computational modeling of recovery \cite{liu2020recovery,hassan2020integrated,monsalve2019data,cassottana2019modeling,guidotti2019integration}, most models are often viewed as resource-intensive black-box approaches and not utilized by communities on the ground. 

The NIST Guide defines time to recovery of function as “a measure of how long it takes before a building or infrastructure system is functioning” and “uses time to recovery of function as the primary metric for community resilience.” This echoes the widely-recognized importance of characterizing disaster recovery for assessing community resilience \cite{Bruneau2003, Chang2010, Cimellaro2010, barabadi2018post}. As the quote by Lord Kelvin says “if you cannot measure it, you cannot improve it,” the lack of rigorous and sound estimation methods for recovery time impedes the measurable progress of resilience improvement. 

\begin{figure}[h]
{\centering
\includegraphics[width=4in]{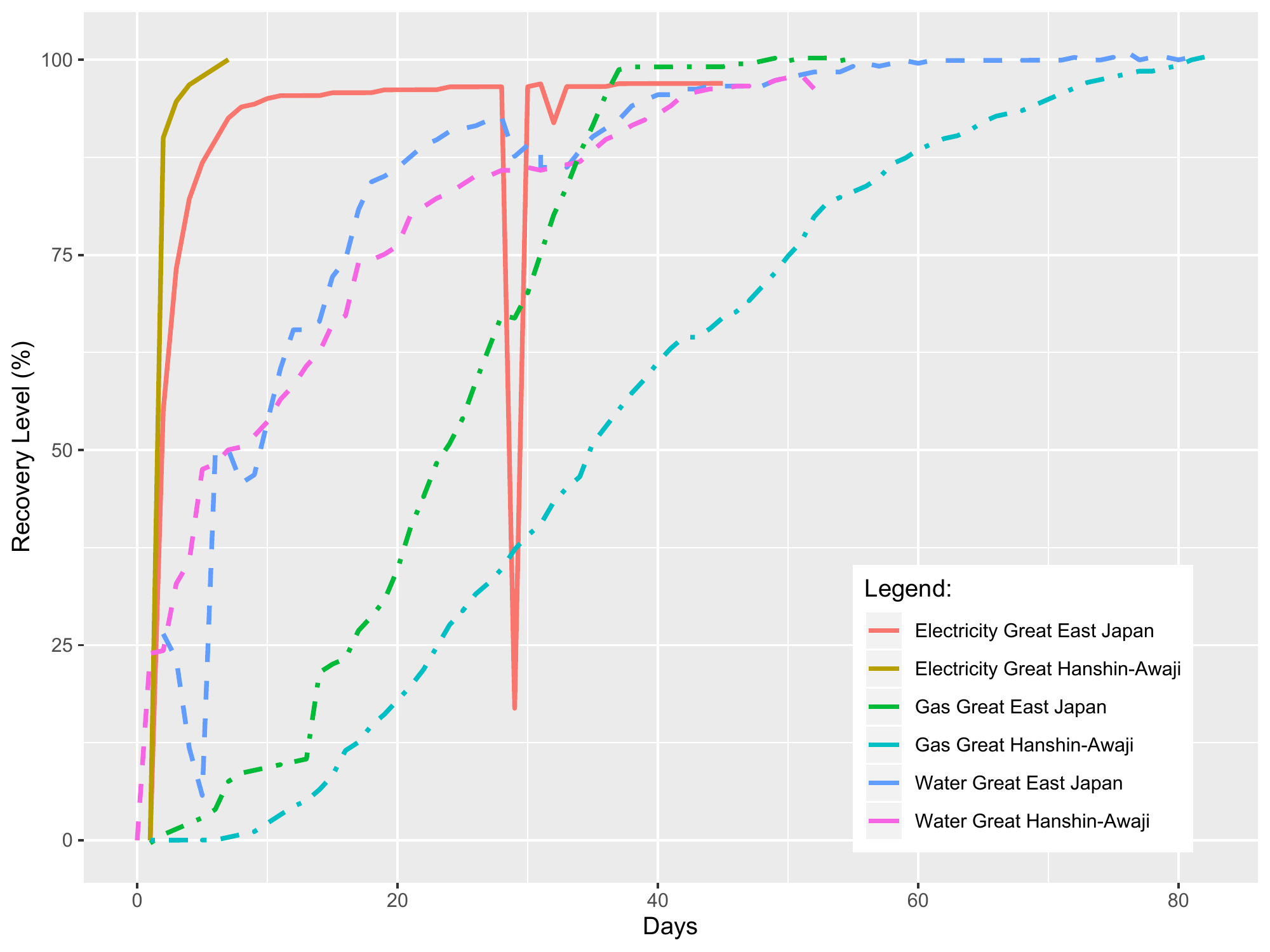}
\\
\caption{\small{Empirical restoration curves of the 1995 Great Hanshin-Awaji Earthquake Disaster and the 2011 Great East Japan Earthquake Disaster.}}
\label{fig:recovery_curves}
}
\end{figure}

To address the above needs and constraints, this paper proposes a statistical framework to estimate infrastructure recovery curves (e.g., see Figure~\ref{fig:recovery_curves}) for a hazard scenario using a combination of expert elicitation and Gaussian process regression (GPR). The two methods complement each other to provide satisfactory solutions to this problem. Estimates gathered from experts will provide initial guidelines on how long it will take for a particular infrastructure to recover to some intermediate functionality levels. GPR will then use these estimates to predict the full recovery curve while capturing potential uncertainty in its prediction, as well as the uncertainty in the experts' estimates. GPR is also flexible enough to enforce important constraints on its predictions to allow the predicted curve to follow the physical behaviour of the actual recovery curve (e.g., monotonically increasing and bounded between 0 and 100\%). The framework aims to be extensible to various types of infrastructure, while being intuitive and easy to be interpreted by the stakeholders.

While more data would generally yield a more accurate estimate, there is a practical limitation on collecting expert-elicited data. We study how to balance between the cost of collecting data from expert elicitation and the estimation accuracy of GPR. We consider multiple expert elicitation schemes to identify the best way to estimate the recovery curve with a reasonable cognitive burden on experts while maintaining good estimation accuracy. 

We simulate expert-elicited data by randomly generating expert estimates, 
which are assumed to be generally close to the empirical recovery curve observed in a case study event. We evaluate the proposed estimation method based on different empirical recovery curves from different prefectures and infrastructures after the 1995 Great Hanshin-Awaji Earthquake and the 2011 Great East Japan Earthquake \cite{nojima2012restoration}.

The rest of this paper is organized as follows.  Section~\ref{sec:background} briefly reviews relevant literature on expert elicitation and GPR.  Section~\ref{sec:method} presents the proposed estimation methodology. 
Section~\ref{sec:num_ex} shows the performance of this method through extensive numerical studies and sensitivity analyses. 
Section~\ref{sec:conclusion} draws insights for potential users of this method and concludes the paper. 

\section{Background}\label{sec:background}
\subsection{Expert elicitation for disaster recovery estimation}
Participatory methods, especially expert elicitation, have been used extensively in disaster research especially in the areas where empirical data are scarce \cite{Miles2006, Miles2011}. 
The study in \cite{Chang2014} elicits from experts infrastructure recovery estimates (at 0 hours, 72 hours, and 2 weeks from a hypothetical event) and qualitative inter-dependencies between those infrastructures. However, the study limits itself to short-term restoration and does not factor uncertainties into the recovery time estimation.

Expert elicitation itself is a well-established research domain \cite{cooke1991experts, gordon2011calculating}. One of the most well-known elicitation approaches is the Delphi method \cite{dalkey1969delphi, brown1968delphi} characterized by its iterative, anonymous approach for developing consensus among experts. This method has been used widely in governments and industries \cite{dalal2011expertlens, dalkey1971comparison}. Another approach is the Cooke Classical Model \cite{cooke1991experts, CLEMEN2008}, also known as Cooke’s method, which is one of the most established methods in expert elicitation literature. This method uses calibration questions, for which true values are known to the facilitator, to measure both accuracy and informative-ness of an individual expert’s judgement. These performance measurements, called calibration score and information score, respectively, are used as weights for aggregating multiple experts’ judgements. Although developing calibration questions requires extra efforts, this performance-based weighting scheme has empirically proven effective \cite{cooke2008tu} and represents the state-of-the-art among various weighting schemes \cite{aspinall2013quantifying, clemen1987calibrating, cooke2008performance}. In this paper, we propose to elicit data from the expert panel using both Delphi and Cooke's methods. The Delphi method is used to estimate a crucial quantity 
that needs a consensus across experts. The Cooke's method is used to aggregate recovery estimates across experts according to performance-based weights. 

Although many studies elicit point estimates or probability distributions from experts, there are only a few studies on eliciting functions (e.g., recovery curve) from experts \cite{zickfeld2007expert,beccacece2015elicitation,durbach2017expert}. Arguably, the most systematic expert elicitation approach to functional estimation is developed in \cite{jaiswal2014estimating}. This study estimates seismic collapse fragility functions by eliciting quantiles of probability distributions, which encompass uncertainties of both seismic shaking intensity and resulting building collapse, from earthquake-engineering professionals. The reported estimates therein are created by first fitting lognormal distributions to the elicited probability estimates and then aggregating the distributions using Cooke’s method. While this approach using the lognormal distribution (often used to model collapse fragilities) is defensible for this study, generalizing the approach to other functional estimation (especially recovery time estimation) has a major drawback. Using a parametric distribution like lognormal is too restrictive to reflect the uncertainties underlying the complex recovery processes being modeled. Thus, this study uses GPR, which allows us to nonparametrically model recovery curves and the associated uncertainties.

Integration of expert judgements and empirical data is briefly mentioned in the NIST Guide \cite{nist2016}, but no specific guideline is provided on the integration. The Oregon Resilience Plan \cite{osspac2013oregon} was the only resilience planning initiative that explicitly used both expert judgements and past event data, but the estimation process was still ad-hoc. Currently, to our best knowledge, there is no systematic statistical inference method being used for expert-based recovery time estimation in practice. This gap inspired us to develop the proposed method. 

\subsection{Gaussian process regression}
Gaussian Process (GP) is a nonparametric model that offers the flexibility to model a stochastic process. It has been used successfully in many applications, such as engineering, physics, biology, economics, or other fields, in both regression and classification problems \cite{Nickisch2008, murphy2012machine, Golchi2015, Riihimaki2010}. It specifies a prior distribution over function spaces, where the relationships over data are encoded in the covariance functions of multivariate Gaussian distributions. Once the input data is available, GP can model the posterior over function spaces. The covariance will determine properties or constraints of the process, such as characteristic length scale, smoothness, or variance \cite{Rasmussen2006gaussian}. In this study, we are estimating recovery curves, so we will focus on Gaussian process regression (GPR). 

\begin{figure}[h]
{\centering
\subfigure[GPR with Noise-free observations]{\includegraphics[width=2.3in]{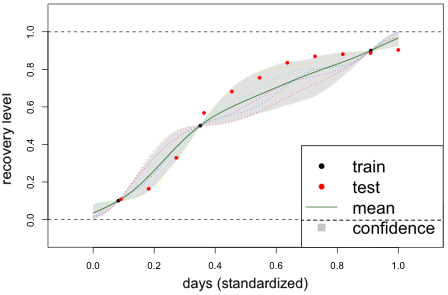}}
\subfigure[GPR with Noisy observations]{\includegraphics[width=2.3in]{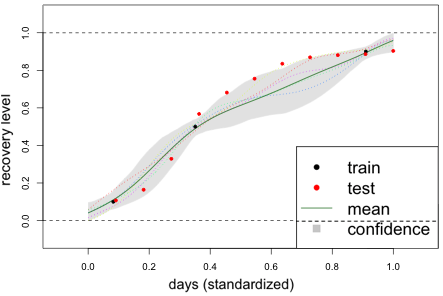}}
\\
\caption{\small{GPR fitting with noise-free and noisy observation for Fukushima prefecture electricity recovery.}}
\label{fig:noise}
}
\end{figure}

Besides its low bias towards any functional form, GPR is also more suitable to our task than other parametric methods. It can capture both the uncertainty in the region where training data is not available and the variability in the training data itself. As a well-known issue in judgement-based forecasting, no matter how rigorous the elicitation process is, the results still depend on the experts' ability to estimate the quantity of interest. 
Because the expert estimates are noisy, GPR will capture the variability as an extra source of uncertainty during the inference step. Figure~\ref{fig:noise} shows two different ways to fit GPR to estimate a recovery curve, with or without noise in the training data.  

\begin{figure}[h]
{\centering
\subfigure[GPR with monotonicity constraint only.]{\includegraphics[width=2.2in]{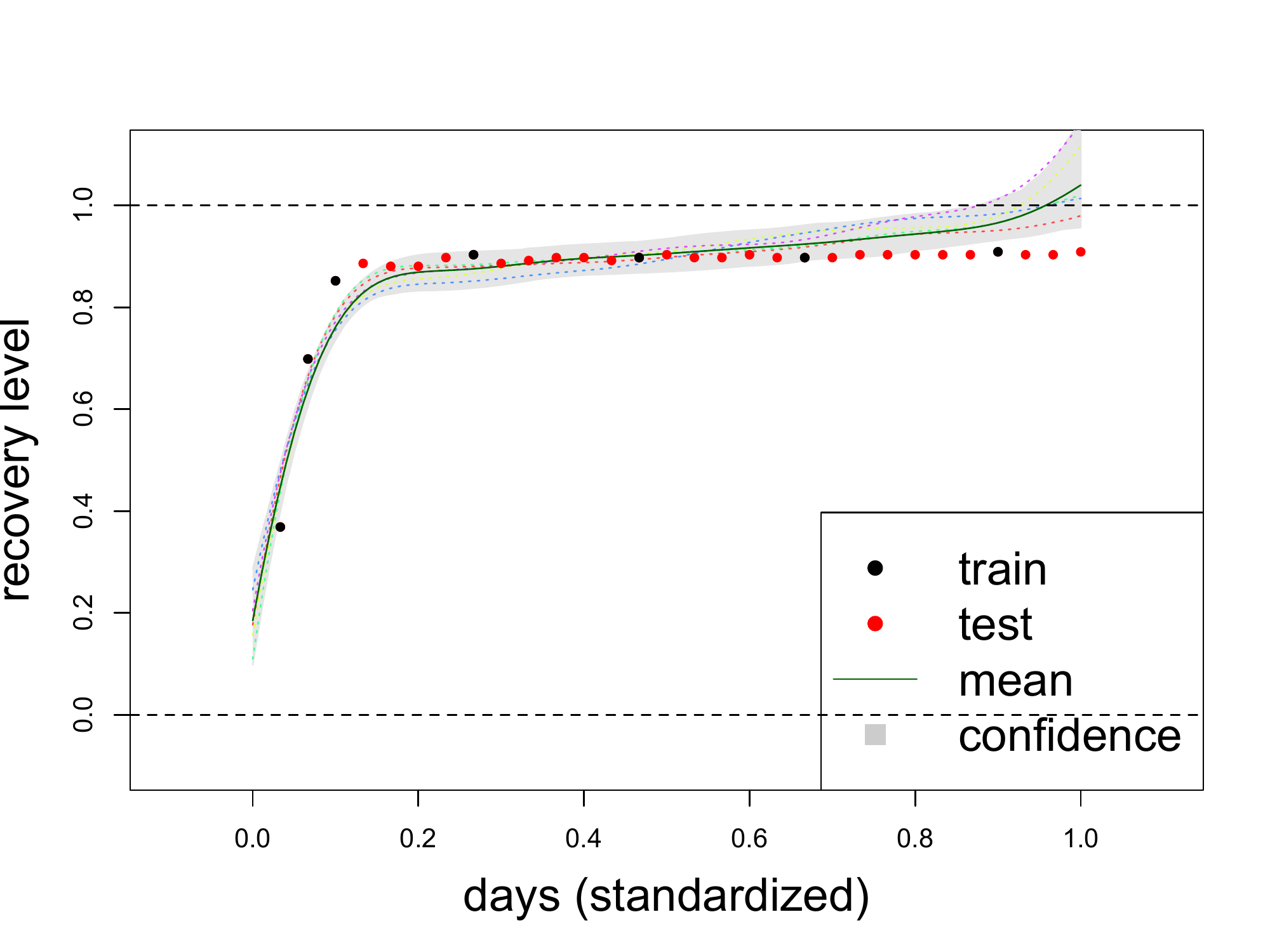}}
\subfigure[GPR with boundedness constraint only.]{\includegraphics[width=2.2in]{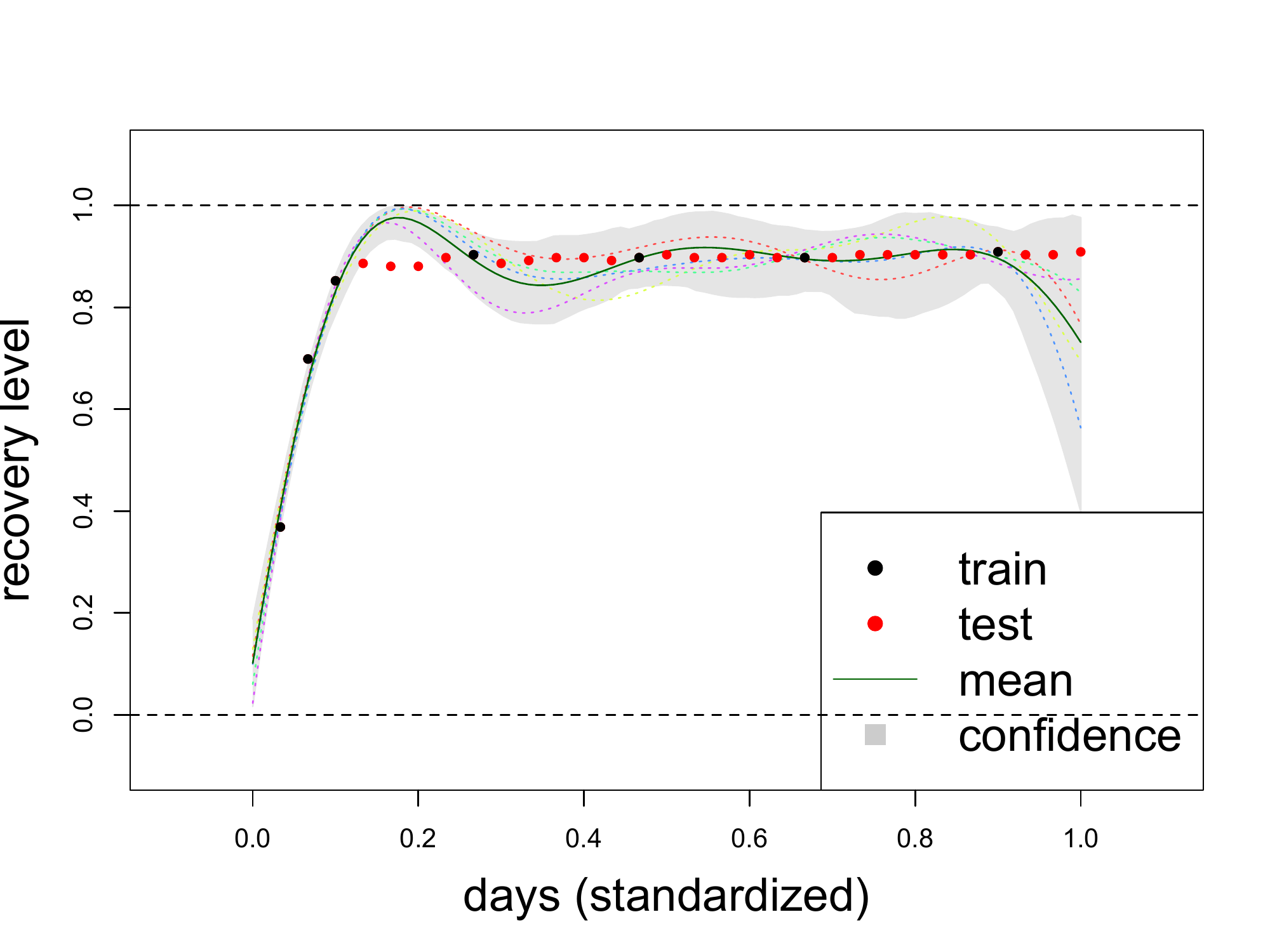}}
\subfigure[GPR with both monotonicity and boundedness constraints.]{\includegraphics[width=2.2in]{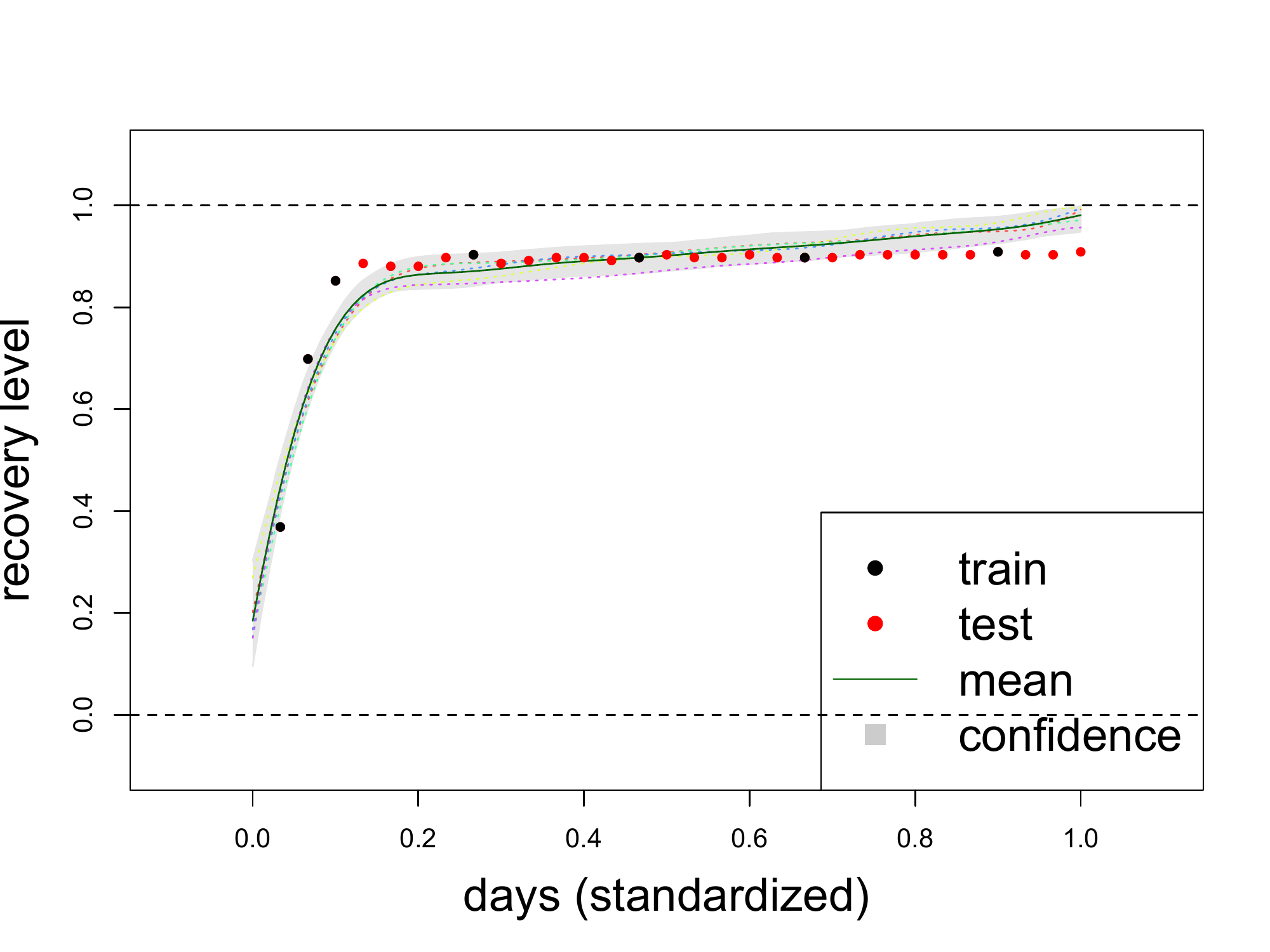}}
\\
\caption{\small{Different GPR constraints for the Fukushima prefecture electricity event.}}
\label{fig:lineq}
}
\end{figure}

Due to the physical nature of the recovery curve, we also need to impose some constraints on the GPR model. First, the functionality level should be between 0\% and 100\%. Therefore, we will bound the prediction of the GPR model to be strictly between 0 and 1. Second, although it is possible that functionality level may temporarily decrease in reality (e.g., due to an aftershock), it should generally increase over time. Hence, to capture this behaviour and reduce the prediction error, we also enforce that the curve is monotonically increasing with respect to time. Montonicity and boundedness are the linear inequality constraints actively researched in the GP framework \cite{Riihimaki2010, Maatouk2017, Maatouk2017a, Lopez-Lopera2018, Lopez-Lopera2019}. In Figure~\ref{fig:lineq}, we show the effects of imposing only monotonicity, only [0,1] boundedness, and both constraints in the model for the Fukushima prefecture electricity recovery using the R package \texttt{lineqGPR} \cite {Lopez-Lopera2018, Lopez-Lopera2019}. It is helpful to have both constraints in the model. Otherwise, the model may behave in contrast to the expected physical behaviour of infrastructure recovery. In addition, the constraints will help to reduce the variance of the prediction. However, imposing these constraints may be potentially too rigid to capture the flat region near 0\% and 90\% of recovery. We can alleviate this issue by eliciting the boundary points so that the GPR is only interpolating between the elicited data points. Furthermore, the recovery curve can be constructed up to a functional level below 100\% (e.g., 90\%) as suggested by the NIST Guide \cite{nist2016}. We will apply these measures in Section~\ref{sec:num_ex} for the numerical studies. 

\section{Method}\label{sec:method}

\subsection{Consideration in recovery curve estimation}
Our goal in this study is to estimate the infrastructure recovery curve from point estimates given by experts using GPR. The two steps (i.e., expert elicitation and GPR) are not designed independently. We carefully design the whole framework considering the logistical, computational, and theoretical constraints of both steps. The curve is characterized by two dimensions, namely, the recovery level measured in percentage (100\% means the system is fully functional) and the recovery time measured in either days or hours from the disruption. GPR, similarly to other regression methods, is more suitable for interpolation between training data (as opposed to extrapolation). To achieve better performance, it is desired for the expert elicited data to possess two properties. First, it should be as evenly distributed as possible so that the interpolated prediction does not exhibit too much uncertainty. Second, it should cover the boundary values to avoid predicting values beyond the range of the given data. It may seem best from a statistical perspective to elicit the data in both the range of recovery level (e.g 10\%, 30\%, 50\%, 70\%, 90\%) and recovery time (e.g 2.5$D$/10, 5$D$/10, 7.5$D$/10, 9$D$/10), where $D$ is the (estimated) earliest time for the infrastructure to recover to 100\% or another high functional level (e.g. 90\%) depending on what kind of recovery curve we want to construct, to follow the NIST Planning Guide \cite{nist2016}) given by the expert. However, it may not be very intuitive to elicit recovery levels at some certain time, e.g. ``What is the estimated recovery level at day 11 after the event?". Therefore, in our proposed method below and the numerical studies in Section ~\ref{sec:num_ex}, we only elicit the recovery time at some certain recovery levels. This choice is also consistent with the NIST Planning Guide \cite{nist2016}. 

In addition to eliciting recovery times at different recovery levels, we also want to elicit $D$, as introduced above, so that we can normalize the recovery time to be in the range of $[0,1]$ for the following reasons:
\begin{enumerate}
    \item The constructed curve could be more generalizable to future disaster events. If we face another similar events in the future, where similarity is defined by dominant characteristics of the events (e.g., Richter magnitude and earthquake resilience of the area, or Saffir–Simpson scale and hurricane resilience of the area), we can significantly reduce the elicitation effort by either using the existing recovery curves or simply eliciting the earliest full recovery time $D$ and scaling the recovery time based on the particular $D$ values of new events.
    \item It is easier to compare different recovery curves of different natures on the same scale in the range of $[0,1]$. 
    \item We can offer some insights from the shape or pattern (e.g., for hurricane category 1 vs. 5; magnitude 6 vs. 8; power vs. water; urban vs. rural) of recovery, which has formed consensus across many communities, so that other communities lacking the opportunity/resource to conduct extensive elicitation procedure can still use these curves as references of possible recovery trends.
    
    \item In terms of GPR modelling, we want to have both axes in the process to be between $[0,1]$ in the fitting and inference following the implementation in \cite {Lopez-Lopera2018, Lopez-Lopera2019}. The actual unit of recovery time can be easily scaled back to days or hours after the inference procedure. 
\end{enumerate}

\subsection{Challenges in expert data elicitation and modelling}

From the design considerations above, we anticipate some challenges in the elicitation process as follows: 
\begin{enumerate}
    \item Obtaining the earliest time to full recovery $D$: $D$ should be universal across all experts. One way to obtain this is to have an open discussion among experts until they reach a general consensus on how long $D$ should be. Another way is to employ point-based expert elicitation methods, for example in \cite{oakley2010shelf}, to estimate the probability distribution of $D$. Another possibility is to use the individual expert's $D$ value to normalize their own recovery time estimate. 
    \item Obtaining the input noise level $\sigma$: This is required for the statistical modelling process. This can be interpreted as how uncertain the experts' estimates are. 
    We may gather the data and estimate the uncertainty based on their data after the elicitation process. This $\sigma$ will account for both within-expert and across-expert uncertainty. The GPR framework assumes that one type of noise is present in the data, which accounts for all the uncertainty, and that the noise level is constant across all levels of input. In case we want to decompose the uncertainty further, it is more straightforward to estimate the across-expert uncertainty since we have different expert data at each recovery level. However, within-expert uncertainty estimation is tricky. One way to estimate it is through Cooke's method where calibration questions are used to measure the inherent estimation uncertainty. However, one can challenge the underlying assumption that the estimated uncertainty based on calibrating questions remains the same as the uncertainty for main questions. Regardless, it is reasonable to assume that within-expert variability is negligible compared with across-expert variability.   
    \item Obtaining more elicited data: While we may drive down the estimation uncertainty by collecting more data, this would impose more logistical burden to the experts. In addition, the experts may have some cognitive difficulty to distinguish between smaller difference in recovery levels, e.g 10\% and 20\%.
\end{enumerate}

On the other hand, there are also some challenges in the modelling and inference process: 
\begin{enumerate}
    \item If we ask each expert to give an estimate of $D$, it is challenging to determine which $D$ to use and how to normalize the time. 
    \item If we have noise/uncertainty in both dimensions (input and output), it does not follow the conventional GPR framework, in which $y = f(x) + \epsilon$, where $\epsilon$ follows $N(0, \sigma^2)$. To further elaborate this point, in the GPR framework, we assume that the input is fixed, i.e. if we want to predict the recovery time at each functionality level, we fix the functionality level and the prediction of recovery time will exhibit some level of uncertainty. This is consistent with our experiment implementation in Section~\ref{sec:num_ex}.
\end{enumerate}

\subsection{Recovery curve estimation framework}
In this section, we present a few potential elicitation schemes for consideration. Each scheme has its own advantage and disadvantage. 

\noindent\textbf{Scheme 1}: Maximum elicitation on two dimensions. 
\begin{enumerate}
    \item Ask each expert for the earliest time to full recovery $D$, recovery times at fixed functionality levels (10\%, 30\%, 50\%, 70\%, 90\%), and functional levels at fixed recovery times (e.g 2.5$D$/10, 5$D$/10, 7.5$D$/10, 9$D$/10).
    \item Use the sample mean/median (across experts) of all elicited data as the training data, with Cooke's method weighting if necessary.
    \item Use all estimates (across experts) at each level to estimate the noise level. 
    \item Fit GPR and construct a recovery curve with its estimation uncertainty.
\end{enumerate}

\underline{Advantage}: Full range of data over both dimensions. Impose less burden in elicitation logistics than scheme 2. Elicitation can finish in one stage. 

\underline{Disadvantage}: Uncertainty in $D$ estimation can lead to erroneous and high uncertainty in prediction. Furthermore, as mentioned above, the GPR framework assumes one dimension as fixed input. Eliciting in both dimensions violates this assumption. 

\noindent\textbf{Scheme 2}: Two-stage elicitation. The earliest full recovery time $D$ will be iteratively discussed among the experts until reaching consensus. 
\begin{enumerate}
    \item[]\hspace{-1.5em}Stage 1:
    \item Ask each expert for the earliest time to full recovery $D$.
    \item Show all the experts the (range of) elicited $D$ values.
    \item Ask experts to revise their $D$ estimate until reaching agreement. 
    \item[]\hspace{-1.5em}Stage 2: 
    \item Ask each expert for a full range of fixed recovery level (10\%, 30\%, 50\%, 70\%, 90\%) and recovery time (e.g 2.5$D$/10, 5$D$/10, 7.5$D$/10, 9$D$/10), with Cooke's method weighting if necessary.
    \item Use the mean/median of across-expert estimates as the training data.
    \item Use all elicited data to estimate the noise level.
    \item Fit GPR and construct a recovery curve with its estimation uncertainty.
\end{enumerate}

\underline{Advantage}: Full range of data over both dimensions. Reduce uncertainty in the earliest full recovery time $D$.  

\underline{Disadvantage}: Two-stage elicitation will require more effort from the expert. 

\noindent\textbf{Scheme 3}: Using either scheme 1 or scheme 2 but with smaller elicited data (e.g., 3 points for each dimension) 

\underline{Advantage}: Less burden on the expert. 

\underline{Disadvantage}: May result in a sub-optimal fit and prediction. 

\noindent\textbf{Scheme 4}: Elicitation on only one dimension, fixing recovery levels and ask for recovery times. 
\begin{enumerate}
    \item Obtain estimate of $D$, following either scheme 1 or scheme 2. (The numerical studies in Section~\ref{sec:num_ex} will use scheme 4 with each expert's estimate of recovery time being normalized by her own estimate of $D$.)
    \item (OPTIONAL) Repeat the process for the 3 scenarios (worst, best, most likely)
\end{enumerate}

\underline{Advantage}: Straightforward in modelling. Simple to interpret and implement. 

\underline{Disadvantage}: Data may be sparse. In some events (e.g the Fukushima electricity recovery in Section~\ref{sec:num_ex}), the recovery is expected to be very fast in the first few hours. The expert may say the recovery is up to 70\% in the first day and 90\% the next day. In this case, the GPR model may not provide much additional values to stakeholders in recovery planning. 





Scheme 1 will speed up the elicitation process since we can elicit on both dimensions. However, the question is whether we need to elicit in both ways (fix the level then elicit the time, and fix the time then elicit the level). Scheme 2 is almost identical to scheme 1, except with the elicitation of the earliest full recovery time $D$ to reach either 100\% or 90\% to normalize the time axis. 
Scheme 3 is a less resource-demanding version of Scheme 1 and 2. In Section~\ref{sec:num_ex}, we will study the optimal number of elicitation levels through sensitivity analysis. Although we can try to elicit in both ways, to be consistent with the GPR framework, we can only use one dimension (either recovery time or recovery level) as input and predict the remaining dimension. Instead of spending experts’ resources on eliciting in both ways, we can use their effort to elicit more recovery time at higher granularity of functionality level or elicit more scenarios (best, worst, most likely).
In view of the above considerations, we will demonstrate the framework of Scheme 4 in the numerical studies in Section~\ref{sec:num_ex}. 

\section{\textcolor{black}{Numerical Studies}}\label{sec:num_ex}



To demonstrate the performance of the framework, we evaluate it on different empirical recovery curves from different prefectures and infrastructures after the 2011 Great East Japan Earthquake and the 1995 Great Hanshin-Awaji Earthquake. The framework is designed to be applied where an expert elicitation workshop is run in conjunction with statistical modelling. For demonstration purpose in this paper, we will simulate the expert opinion. 
Assuming that the experts are capable of estimating the true recovery curve with a reasonable accuracy, we use the entire available empirical data (such as those in Figure~\ref{fig:recovery_curves}) to fit the polynomial regression model as a surrogate to the expert opinion. The simulated expert can be queried for recovery time given a functionality level and vice versa. In Scheme 4, we provide a functionality level as an input to the simulated expert and obtain the recovery time estimate as the output. Each expert can be modelled using Eq.~\eqref{eq:linearmodel}: 
\begin{equation} \label{eq:linearmodel}
days = f_{poly}(recovery) + \epsilon_{days} , 
\end{equation}
where $days$ is the number of days from the beginning of disruption (e.g., earthquake, hurricane landfall), $recovery$ is the functionality level of the system that is recovering, $f_{poly}$ is the polynomial regression function, and $\epsilon_{days}$ is assumed to follow a normal distribution with mean 0 and variance 
$\sigma^2$ that captures the estimation variability. 

\begin{figure}[h]
{\centering
\subfigure[Polynomial fit in the log scale of the days/recover time]{\includegraphics[width=2.3in]{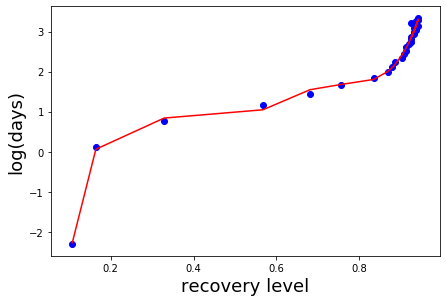}}
\subfigure[Predicting recovery time in the original scale]{\includegraphics[width=2.3in]{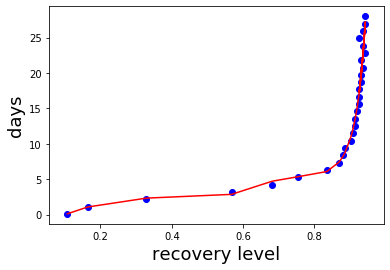}}
\\
\caption{\small{Expert simulated model is built based on all the available data of a past event using polynomial regression, which will allow the sampling from the curve will be very close to the actual values. The model represents an average prediction across multiple replications and multiple expert. In other words, if we have infinite amount of experts, we assume that their average prediction will converge to the fitting curve or the actual data.}}
\label{fig:expert}
}
\end{figure}

However, it is better for the model to utilize the fact that the output (i.e., recovery time) is always positive by taking log transformation on the output variable. Thus, the expert model in Eq.~\eqref{eq:linearmodel} becomes 
\begin{equation} \label{eq:logmodel}
\log(days) = g_{poly}(recovery) + \epsilon_{log(days)}
\end{equation}
The fitted polynomial regression function $g_{poly}$ is demonstrated in Figure ~\ref{fig:expert}.

Furthermore, it is desirable to model two distinct sources of the estimation variability $\epsilon_{log(days)}$. Thus, a layer of Gaussian noise $\epsilon_1$ is added to the output to model within-expert variability. A second layer of Gaussian noise  $\epsilon_2$ is then added to model across-expert variability. The two noise terms are additive in the log-transformed model because we model the errors to be multiplicative in the original scale. 
The log transformation will then make the multiplicative errors become additive, to be consistent with the polynomial regression framework. 
The multiplicative errors are intuitive. For example, consider a scenario event that makes the recovery estimation challenging for all experts (i.e., high across-expert variability, $Var(\epsilon_2)$). Then, the individual expert's large uncertainty perhaps due to lack of experience (i.e., high within-expert variability, $Var(\epsilon_2)$) will \textit{amplify} the effect of the challenging estimation problem, thus resulting in highly variable recovery time estimates. 
In summary, the elicited recovery time estimates are simulated using
\begin{equation} \label{eq:simulated}
days_{simulated} = \exp(g_{poly}(recovery) + \epsilon_1 + \epsilon_2).
\end{equation}
As an implementation note, due to the randomness from $\epsilon_1$ and $\epsilon_2$, sometimes the sequence of simulated expert's estimates could be non-monotonic. How likely it happens depends on the variance of the errors. Since we assume the experts are only providing estimates for a monotonic recovery curve (i.e., no deterioration of infrastructure functionality in the midst of recovery, for example, due to aftershocks), they will only provide monotonically increasing recovery time estimates with respect to the functionality level. In our simulation, to ensure that the simulated recovery estimates satisfy this assumption, we reject the non-monotonic estimate paths until a monotonic sequence is generated. 

Using the simulated data, the GPR model with monotonicity and boundedness is fit as follows:
\begin{equation} \label{eq:gpr}
recovery = GPR(days_{simulated}).
\end{equation}

\begin{figure}[!h]
{\centering
\subfigure[Miyagi electricity recovery curve with 95\% confidence interval]{\includegraphics[width=2.3in]{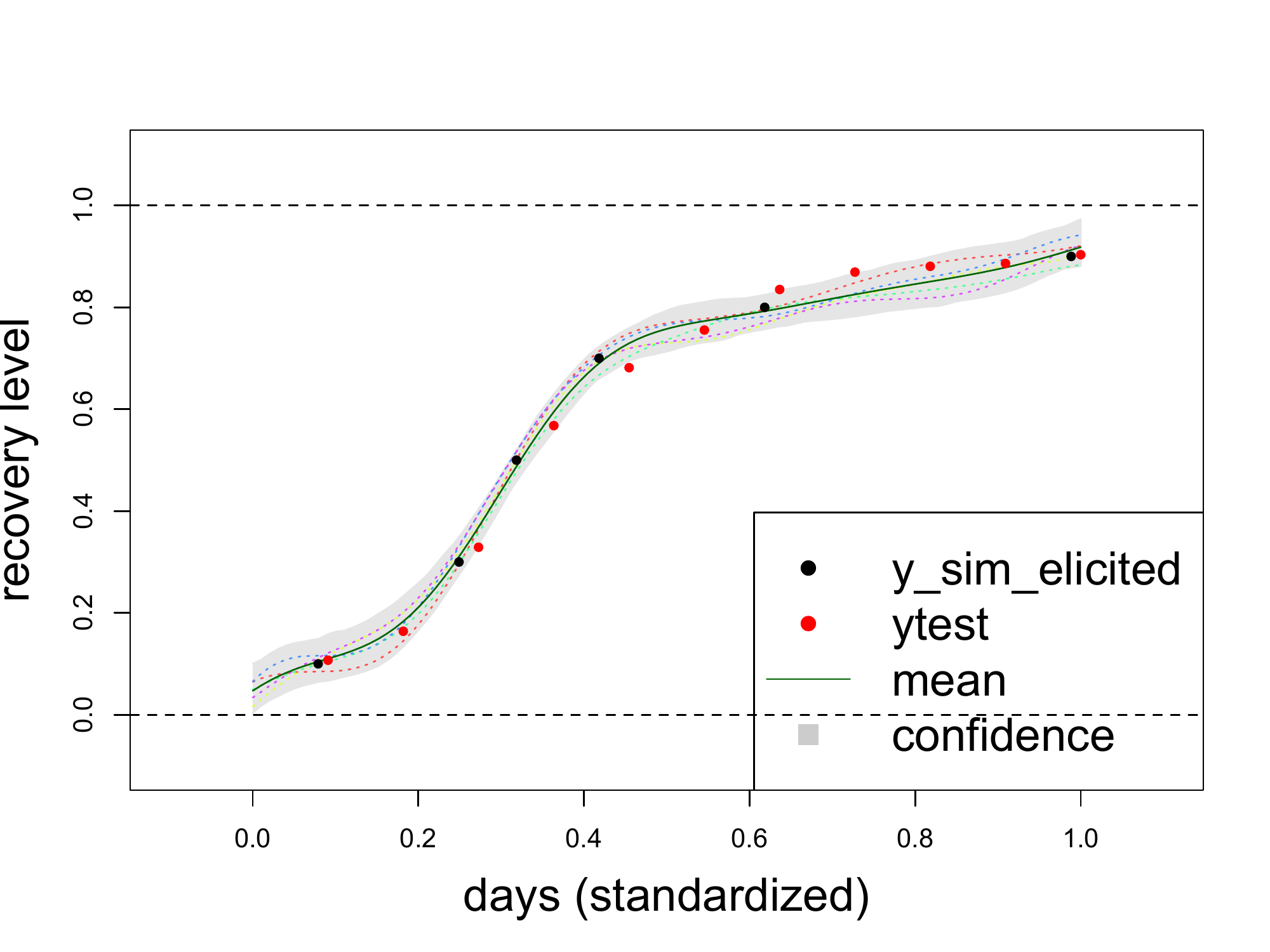}}
\subfigure[Miyagi electricity recovery curve with 10 mean predictions]{\includegraphics[width=2.3in]{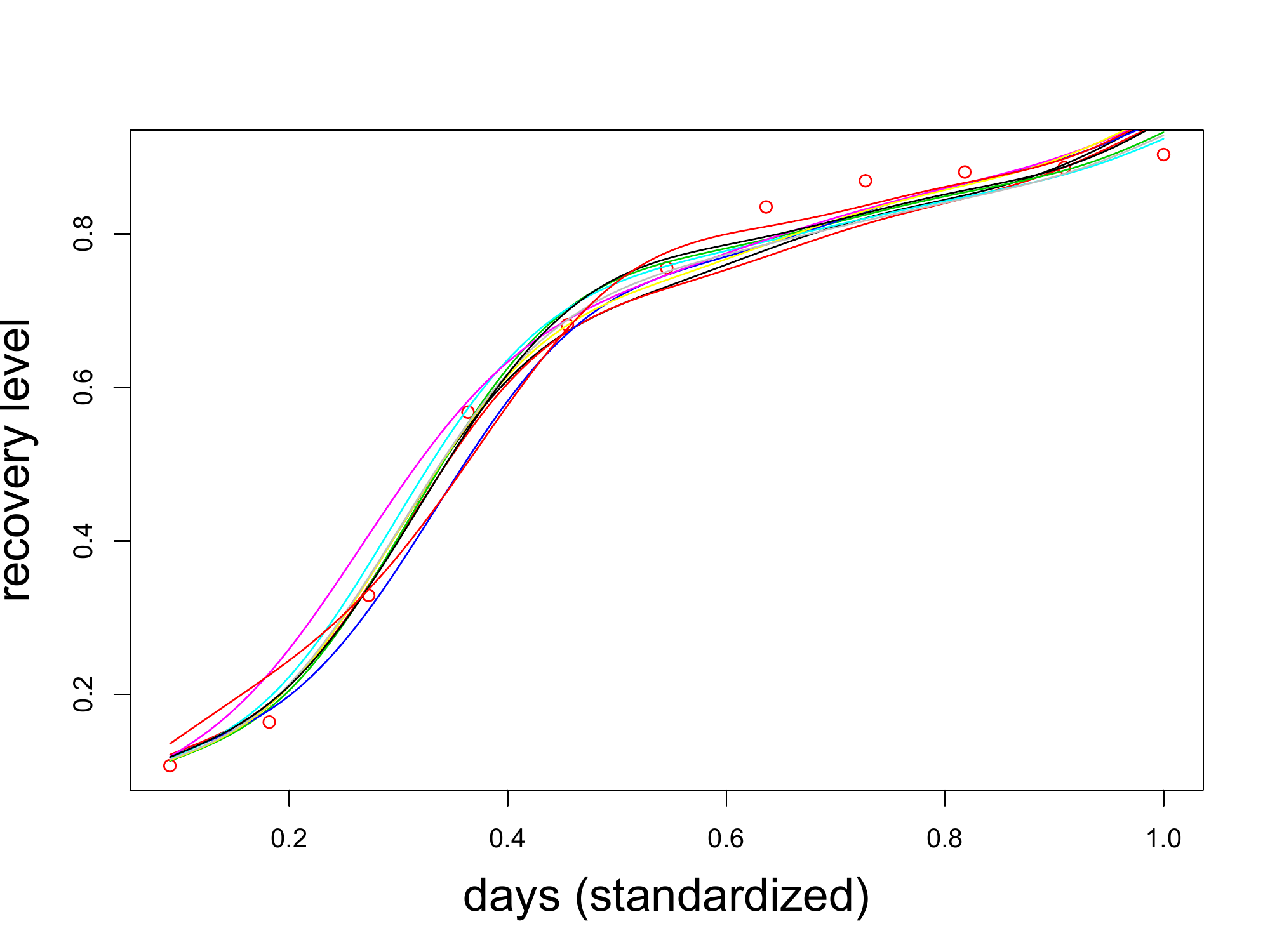}}
\subfigure[Fukushima electricity recovery curve with 95\% confidence interval]{\includegraphics[width=2.3in]{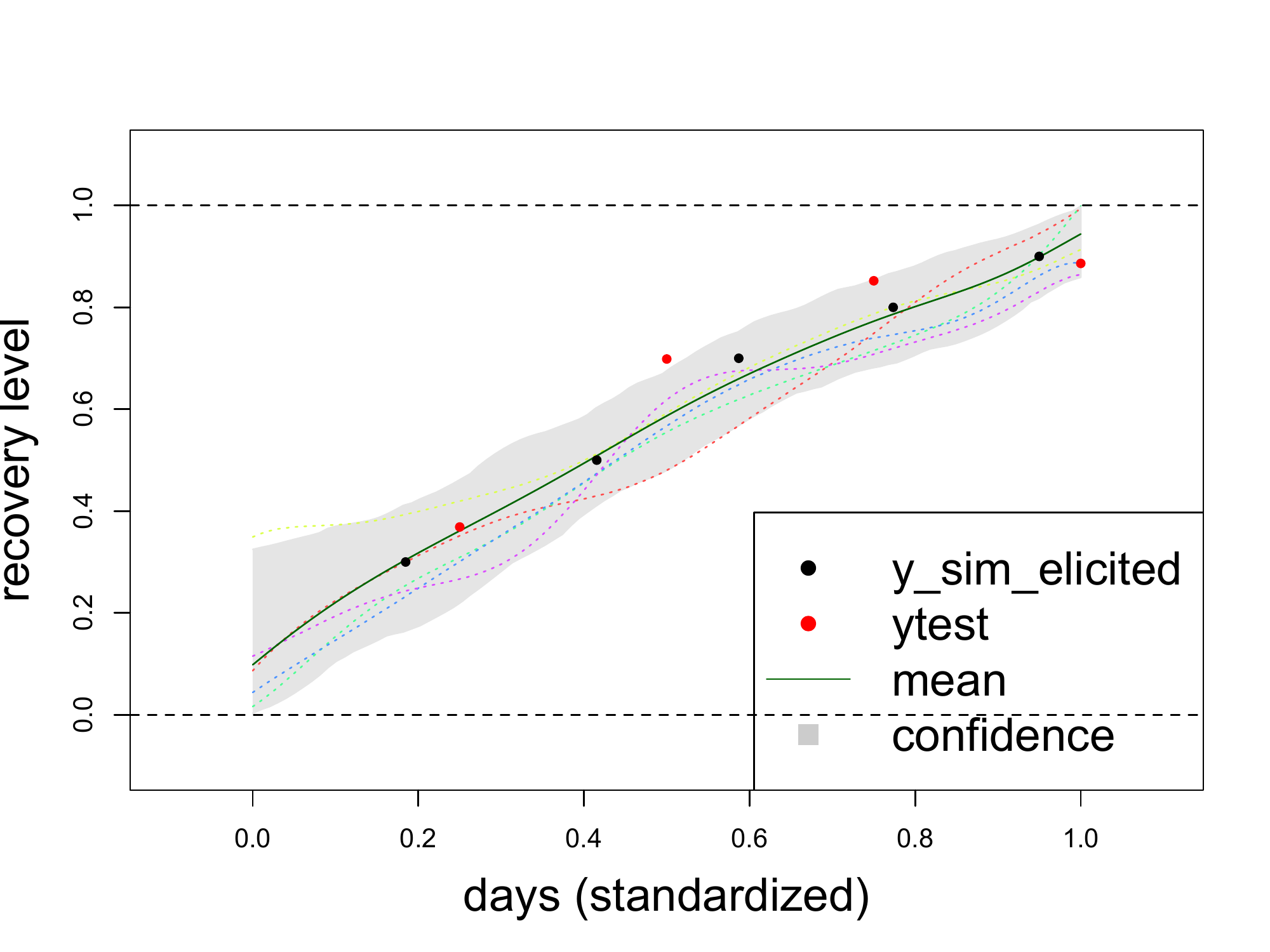}}
\subfigure[Fukushima electricity recovery curve with 10 mean predictions]{\includegraphics[width=2.3in]{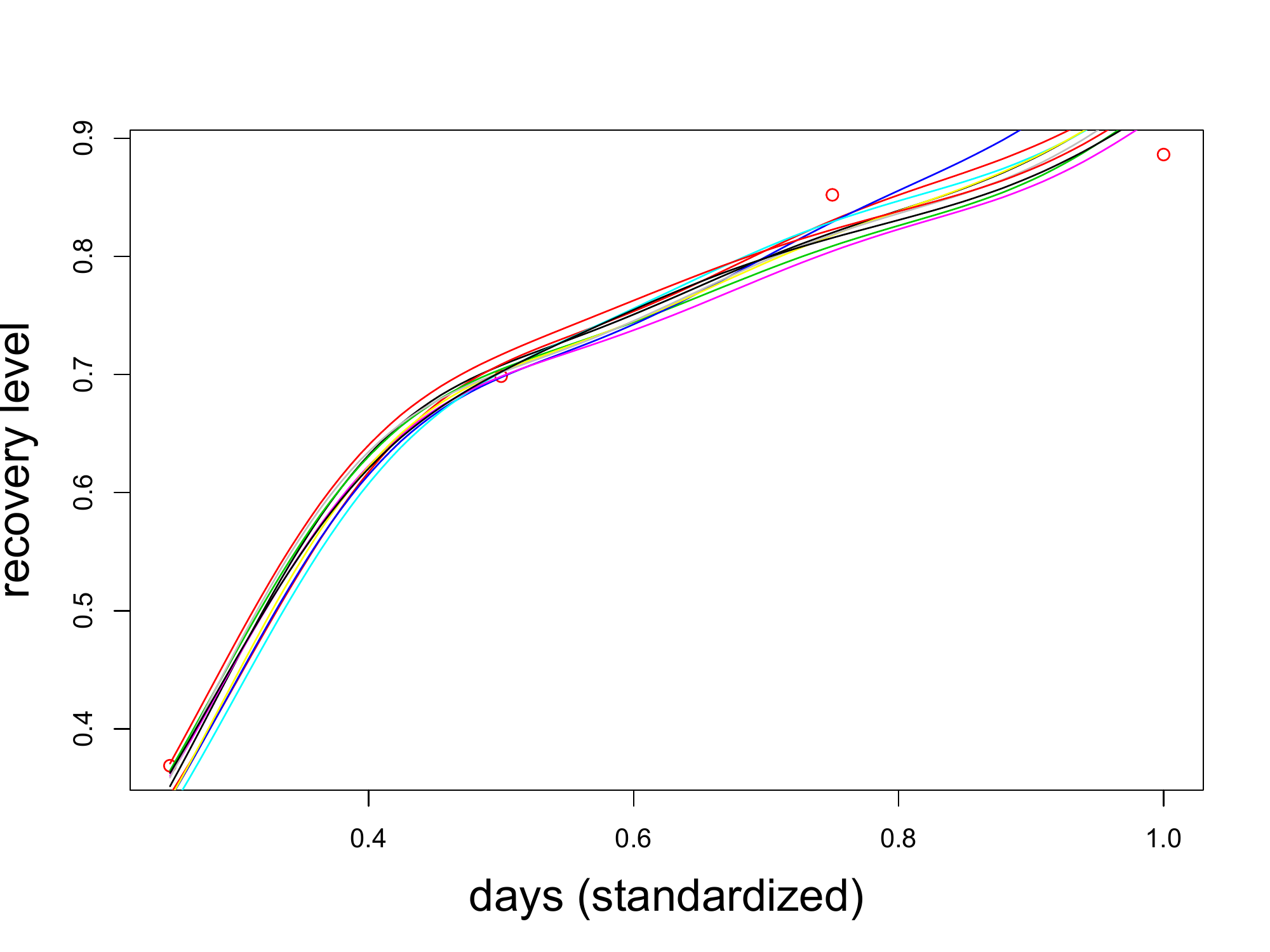}}
\subfigure[Iwate electricity recovery curve with 95\% confidence interval]{\includegraphics[width=2.3in]{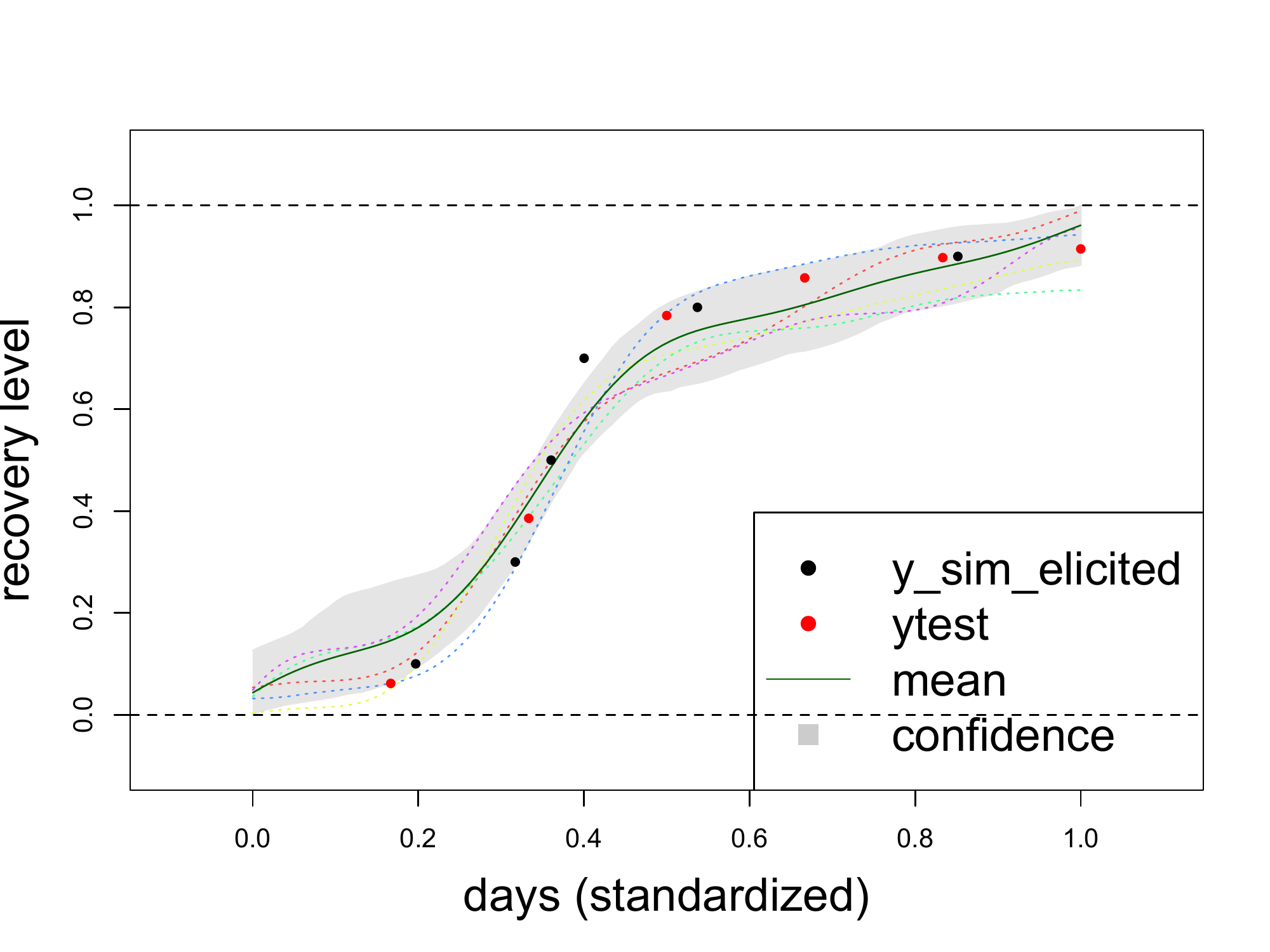}}
\subfigure[Iwate electricity recovery curve with 10 mean predictions]{\includegraphics[width=2.3in]{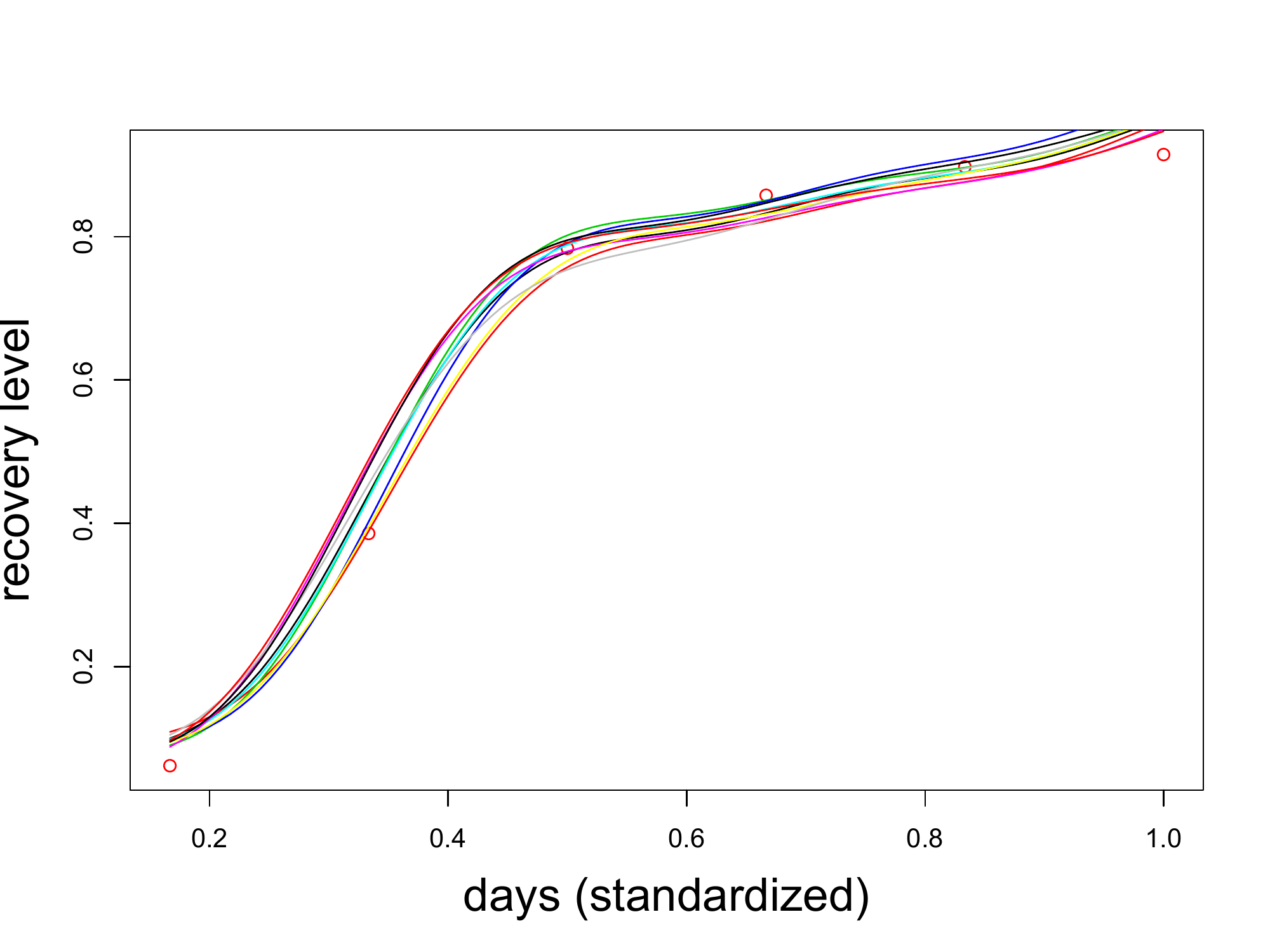}}
\\
\caption{\small{Numerical results on different prefectures (Miyagi, Fukushima, and Iwate). The figures on the left column show the result of GPR model built on one simulated draw of expert opinion. The grey bands show the 95\% confidence interval to capture the uncertainty around the predicted curves. The figures on the right column show different mean predictions based on different simulated draws of expert opinion. In all cases, we simulate the process of elicitation from 5 experts.}}
\label{fig:electric}
}
\end{figure}

Figure ~\ref{fig:electric} shows the performance of this modeling framework on the same infrastructure sector in different prefectures (electricity of Miyagi, Fukushima, and Iwate) and different infrastructures within the same prefecture (water and gas in Great Hanshin). We simulate the process of eliciting from 5 experts, asking for recovery time at 10\%, 30\%, 50\%, 70\%, 90\% functionality levels, averaging their estimates, and construct the GPR curve. 

It is observed that the method is very flexible. In Fukushima electricity recovery, although the actual recovery started at about 40\% in day 1, we can still capture the rest of the recovery curve simply by eliciting from 30\% onward. This translates to some freedom to the experts in actual workshops. They can skip some levels if they think it does not make sense to estimate when they think the recovery actually will happen quickly initially. 

\begin{figure}[h]
{\centering
\subfigure[Great Hanshin water recovery curve with 95\% confidence interval]{\includegraphics[width=2.3in]{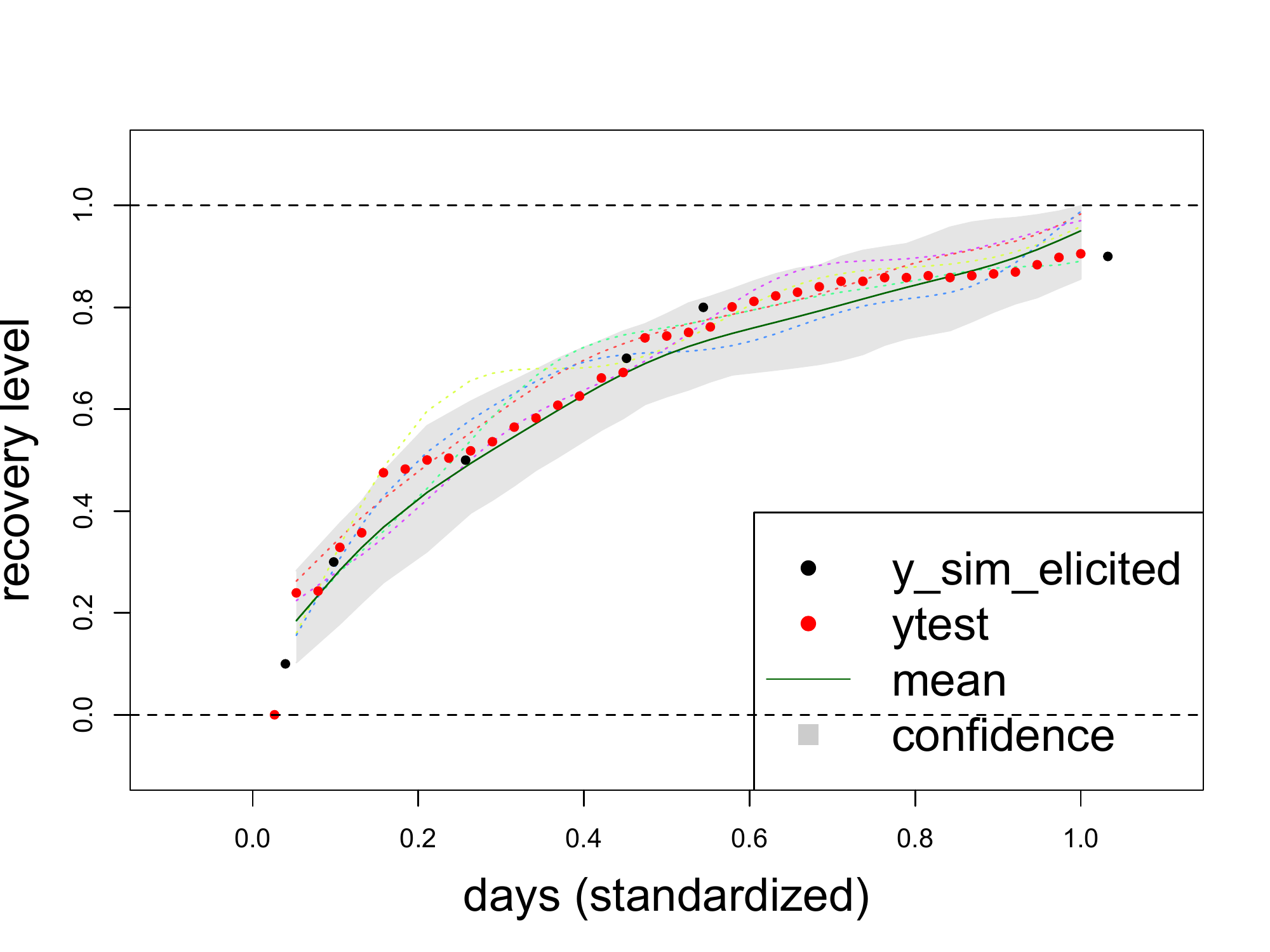}}
\subfigure[Great Hanshin water recovery curve with 10 mean predictions]{\includegraphics[width=2.3in]{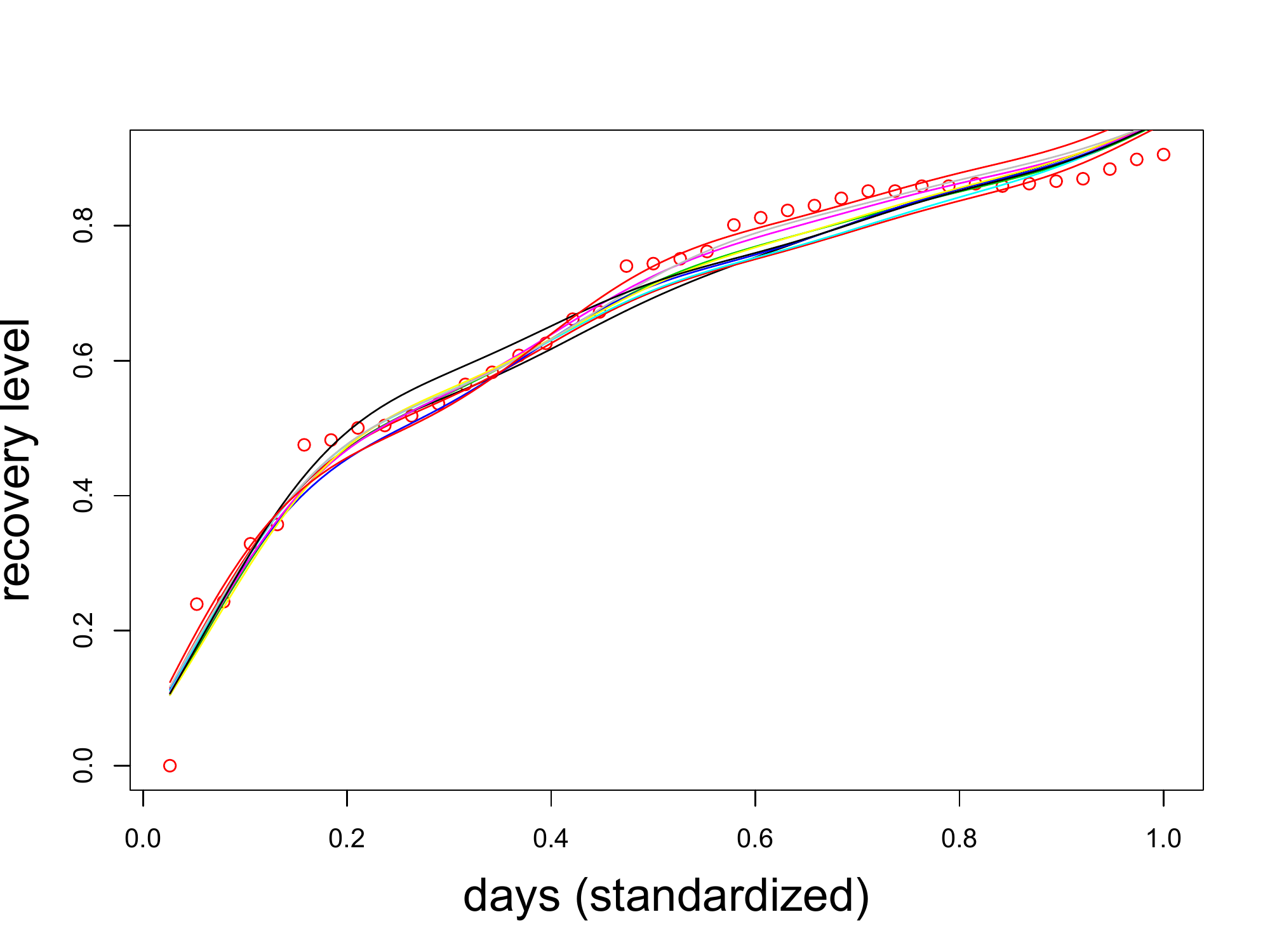}}
\subfigure[Great Hanshin gas recovery curve with 95\% confidence interval]{\includegraphics[width=2.3in]{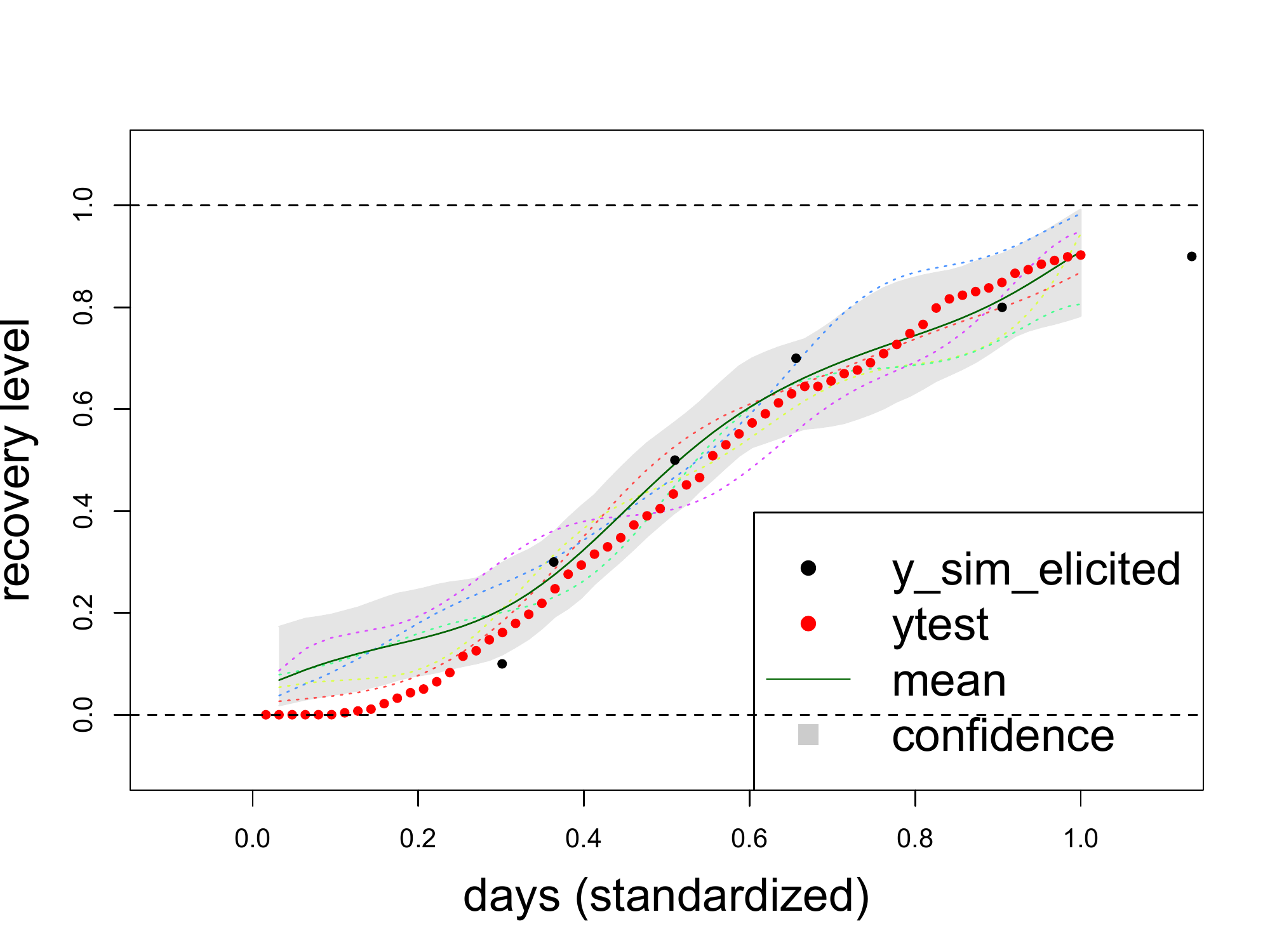}}
\subfigure[Great Hanshin gas recovery curve with 10 mean predictions]{\includegraphics[width=2.3in]{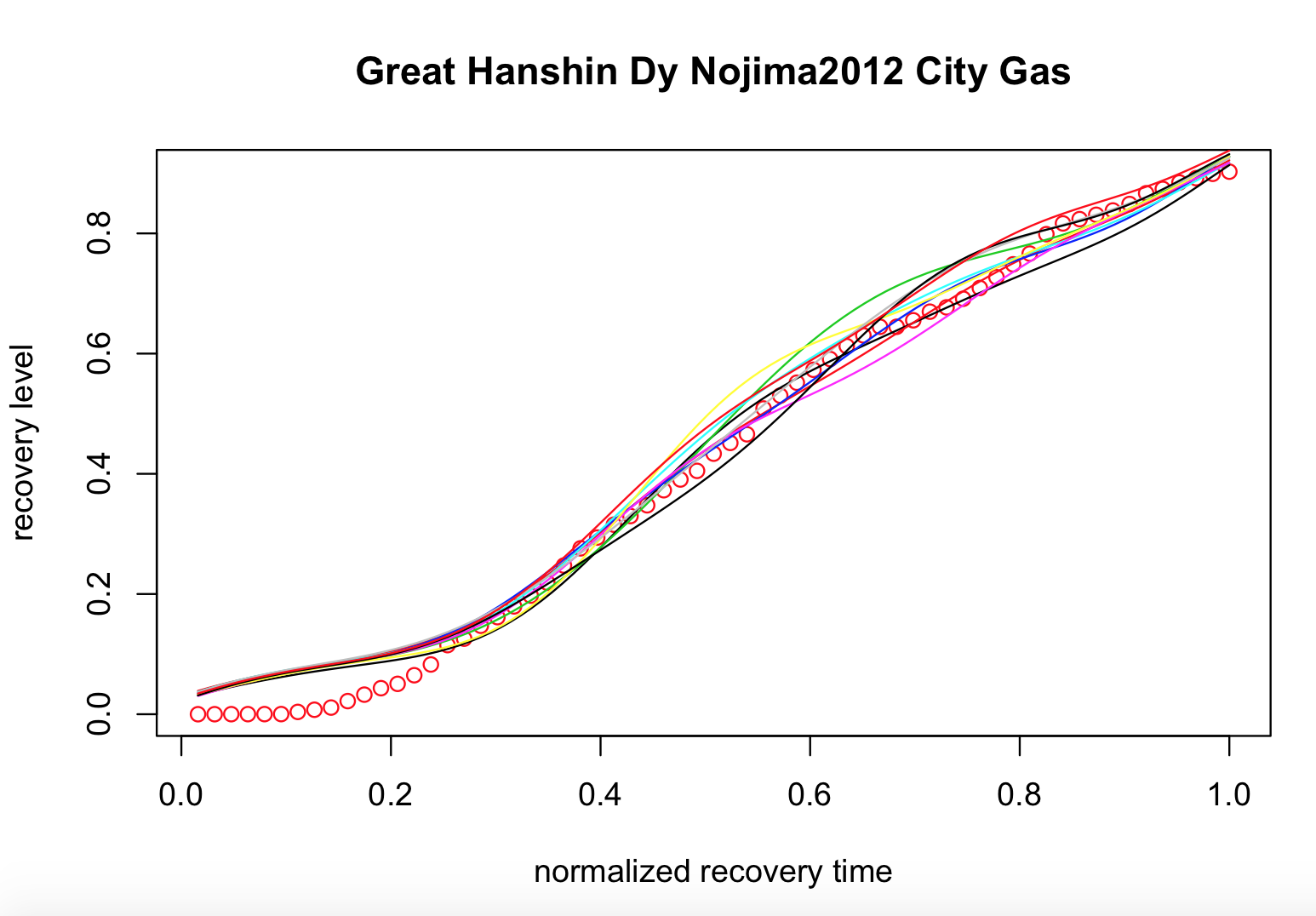}}
\\
\caption{\small{Numerical results on water supply and natural gas infrastructures in Great Hanshin. The data simulates the process of elicitation from 5 experts.}}
\label{fig:utilities}
}
\end{figure}

In Figure ~\ref{fig:utilities}, it may seem that the model does not capture the initial recovery stage (e.g., below 10\% functionality)  very well. This often happens with infrastructures whose recovery tends to follow others, such as gas, which is usually recovered after electricity and water. The model still captures the majority of the recovery curve (between 10\% and 90\%) 
quite well with high confidence. 

\begin{figure}[h]
{\centering
\includegraphics[width=4in]{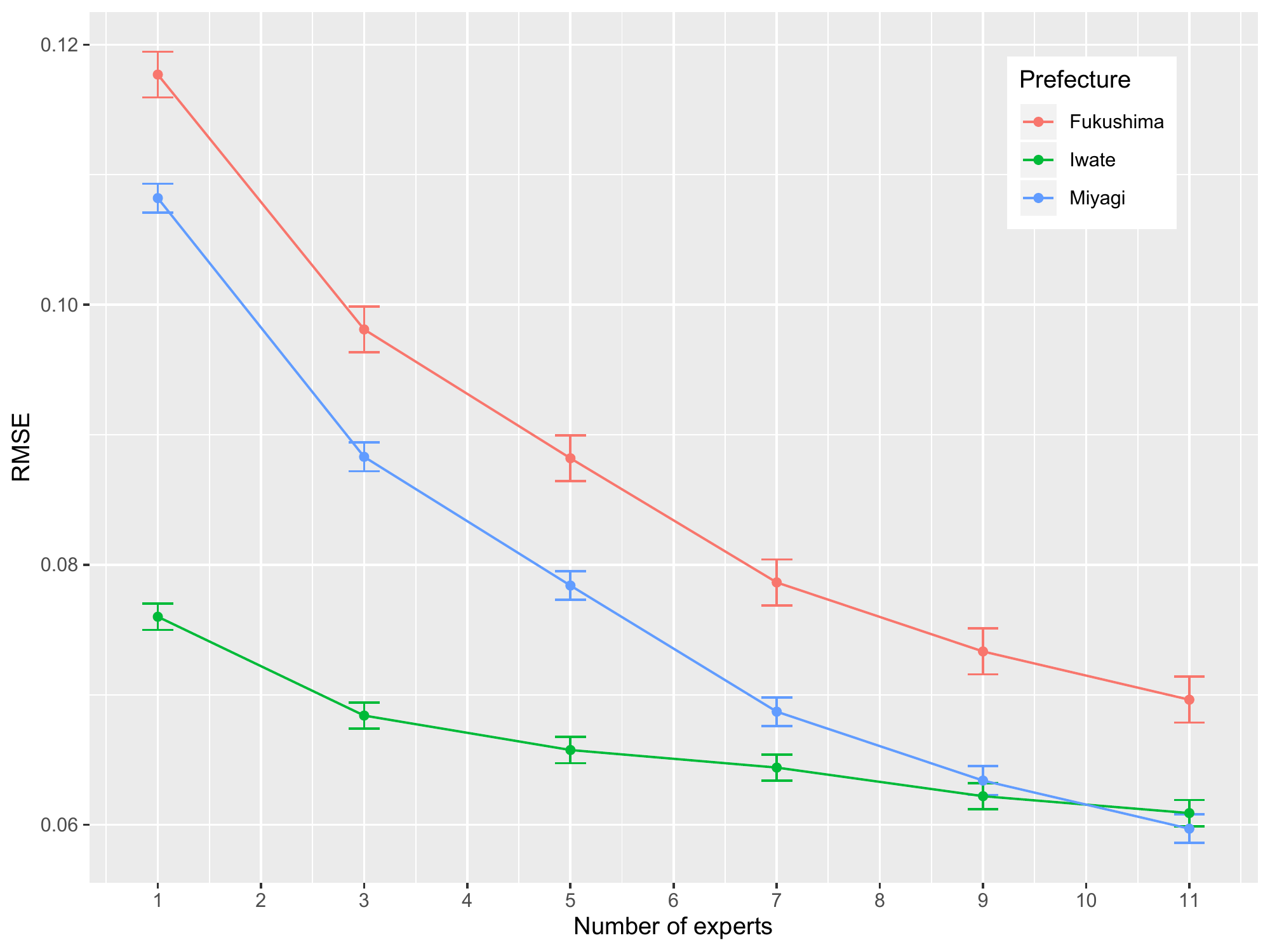}
\\
\caption{\small{The plot shows the performance of the framework for electricity recovery at Fukushima, Miyagi, and Iwate prefactures as a function of the number of experts. The error bar at each number of experts shows the 95\% confidence interval of test RMSE in 100 simulation replications. Although the more number of experts involved in the elicitation process results in better performance, it is observed that there is a diminishing marginal return as the number of experts increases in 2 out of 3 prefactures. The rate of performance gain is fastest when engage from 1 to 3 experts. The rate is slower from 3 to 7 experts. It drops to the slowest rate if we increase from 7 to 11 experts.}}
\label{fig:experts_errorbar}
}
\end{figure}

We also investigate how sensitive the estimation framework is to the number of experts by monitoring the root mean square error (RMSE) of prediction on the available test data (different from the recovery levels elicited from the experts). We first perform simulation to measure the performance in terms of RMSE of the framework with 1, 3, 5, 7, 9, 11 experts based on Miyagi electricity recovery to see if there is an ``elbow" of performance change point to balance the logistics of elicitation and accuracy, as shown in Figure~\ref{fig:experts_errorbar}. In Table ~\ref{tab:expert_num}, we vary the number of simulated experts to be 3, 5, or 10. Given a fixed noise level within and across experts, it seems that the result is quite stable with 5 experts. We acknowledge that in this simulation, all the experts are modelled to exhibit the same level of uncertainty, which is not realistic in practice. In fact, in usability testing experiments in \cite{faulkner2003beyond}, where the participatory performance, involving both expert and novice users, is measured in a group of 5 and beyond, the study shows that some randomly selected group of 5 participants can perform relatively well although the risk is that the performance variance is high. However, in actual workshops, there could be more than 5 experts (among whom, the expertise level is theoretically more consistent than the study in \cite{faulkner2003beyond}), and as long as their opinions converge to some underlying quantity, the estimation still can provide a reasonable recovery curve. 

\begin{figure}[h]
{\centering
\includegraphics[width=4in]{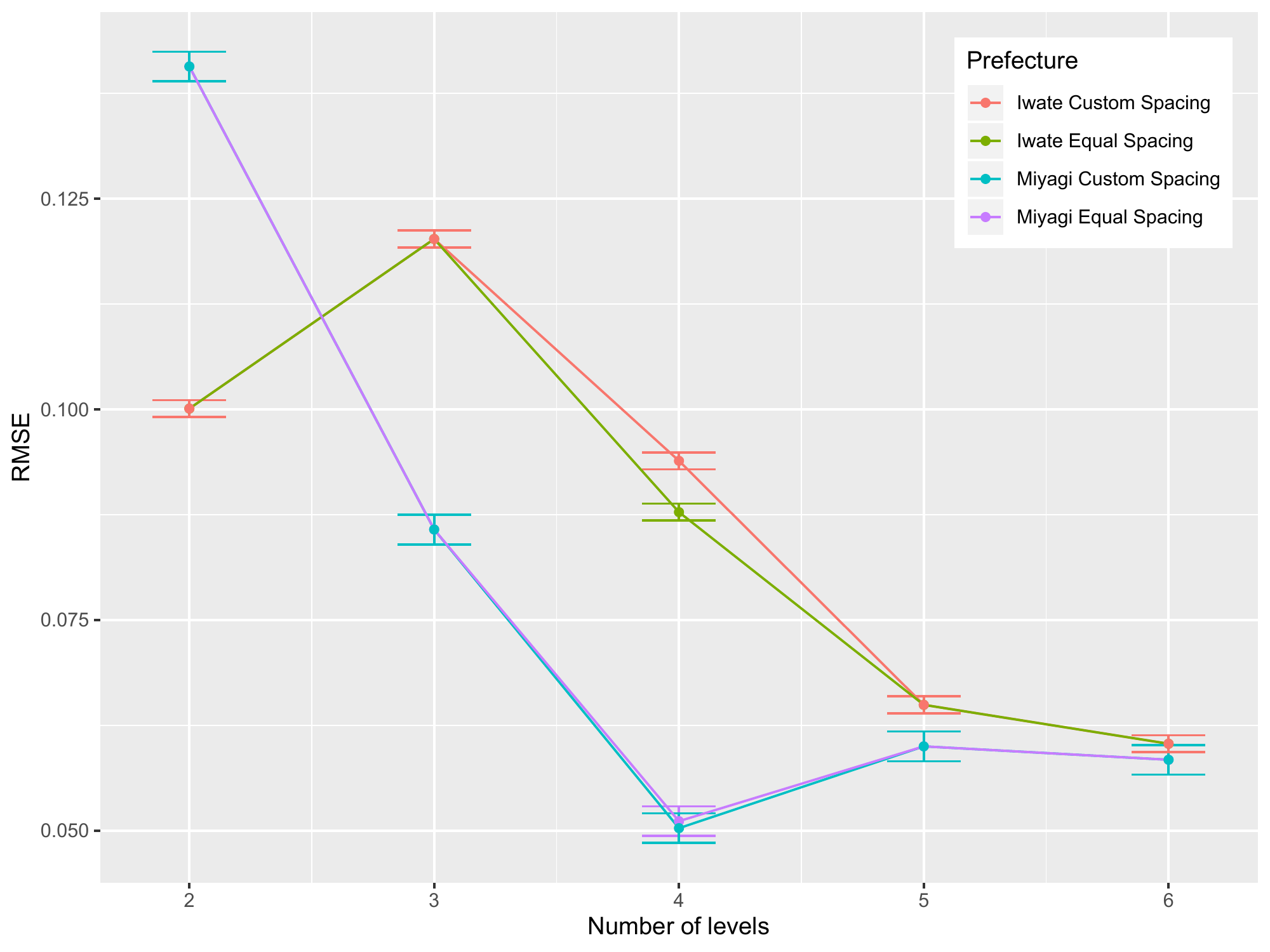}
\\
\caption{\small{The plot shows the performance of the framework in terms of test RMSE in 100 simulation replications for electricity recovery at Miyagi and Iwate prefectures as a function of the number of elicitation levels. We evaluate the performance when eliciting 2, 3, 4, 5, 6 levels from the experts. Custom spacing means we fix the elicited recovery levels at intuitive levels to the experts such as 10\%, 30\%, 50\%, etc, regardless of the number of levels. Equal spacing means we get the levels by equally dividing the range from 10\% to 90\% by the number of levels, which results in some odd levels, such as 10\%, 36.67\%, 63.33\%, 90\% at 4 elicitation levels. The plot shows some general trends that at 4 to 5 elicitation levels, the performance can be satisfactory.}}
\label{fig:levels_errorbar}
}
\end{figure}

We conduct another analysis to measure how the framework performs with different levels of elicitation. Our initial hypothesis is that performance will improve as we elicit more data, which may increase more logistical burden to the expert. The hypothesis is generally confirmed from Figure~\ref{fig:levels_errorbar}. It also does not penalize performance very much to have custom spacing of levels, so we can focus more on asking the experts at more intuitive recovery levels. 

\begin{table}[h]
\centering
{\color{black}\caption{Sensitivity analysis on the framework performance to the number of experts. In this table, the experts are simulated to have equal weights to their estimate, and the simulated noise variance in Eq.~\eqref{eq:simulated} is $Var(\epsilon_1) = Var(\epsilon_2) = 0.1$. Note that the unit for RMSE is the fraction of recovery level. The RMSE presented is the average across 100 simulation replications.}}

\label{tab:expert_num}
{\color{black}\begin{tabular}{ccc}
\hline
Prefacture/ Infrastructure & Number of Experts & RMSE       \\ \hline\hline
Fukushima electricity      & 3  & 0.0637           \\
                        & 5 & 0.0567     \\ 
                        & 10 & 0.0524      \\ 
Miyagi electricity      & 3  & 0.0405          \\
                        & 5 & 0.0340     \\ 
                        & 10 & 0.0297      \\ 
Iwate electricity      & 3  & 0.0457          \\
                        & 5 & 0.0398     \\ 
                        & 10 & 0.0373      \\ 
Great Hanshin water      & 3  & 0.0447          \\
                        & 5 & 0.0352    \\ 
                        & 10 & 0.0334     \\ 
Great Hanshin gas       & 3  & 0.0542         \\
                        & 5 & 0.0516    \\ 
                        & 10 & 0.0497     \\ 
\hline
\end{tabular}}
\end{table}

\begin{table}[h]
\centering
{\color{black}\caption{Sensitivity analysis on the framework performance to the uncertainty in expert estimation ($Var(\epsilon_1), Var(\epsilon_2)$ in Eq.~\eqref{eq:simulated}. In this table, data is simulated from 5 experts for 100 simulation replications.}}
\label{tab:expert_noise}
{\color{black}\begin{tabular}{ccc}
\hline
Prefacture/ Infrastructure & $Var(\epsilon_1), Var(\epsilon_2)$ & RMSE       \\ \hline\hline
Fukushima electricity      & 0.1  & 0.0567          \\
                        & 0.3 & 0.0600    \\ 
                        & 0.5 & 0.0985      \\ 
Miyagi electricity      & 0.1  & 0.0340         \\
                        & 0.3 & 0.0583    \\ 
                        & 0.5 & 0.0811      \\ 
Iwate electricity      & 0.1  & 0.0398          \\
                        & 0.3 & 0.0549     \\ 
                        & 0.5 & 0.0686      \\ 
Great Hanshin water      & 0.1  & 0.0352          \\
                        & 0.3 & 0.0427   \\ 
                        & 0.5 & 0.0546    \\ 
Great Hanshin gas       & 0.1  & 0.0484         \\
                        & 0.3 & 0.0636    \\ 
                        & 0.5 & 0.0916     \\
\hline
\end{tabular}}
\end{table}

\section{Conclusion}\label{sec:conclusion}
We demonstrated in this research a framework to assist the community resilience planning through estimating potential infrastructure recovery curves. The framework combines experts' opinions and Gaussian process regression to unify domain knowledge and uncertainty quantification in the estimated curves. We performed extensive sensitivity analyses to draw insights into various elicitation schemes and the effects of number of experts, number of elicited points, and elicitation levels on the predictive performance. Although the framework was developed for modeling post-event infrastructure recovery, it can be generalized to other recovery modeling, such as for different capitals and services that are important for community resilience \cite{miles2015foundations}. We do not explicitly consider dependencies between infrastructures in this study. Future work may directly model their dependencies to improve the predictive performance and/or reduce the reliance on expert estimates. 

\clearpage

\section*{Acknowledgements}
This work was supported by the National Science Foundation (NSF grant CMMI-1824681).


\bibliography{mybibfile}

\end{document}